\documentclass[%
 reprint,
 amsmath,amssymb,
 aps,
]{revtex4-2}
\usepackage{mathtools}
\usepackage{physics}
\usepackage{braket}
\usepackage{here}
\usepackage[dvipdfmx]{graphicx}
\usepackage{dcolumn}
\usepackage{bm}
\usepackage{newtxtext,newtxmath}
\usepackage{hyperref}
\hypersetup{
setpagesize=false,
bookmarksnumbered=false,%
bookmarksopen=false,%
colorlinks=true,%
linkcolor=blue,
citecolor=blue,
urlcolor=blue,
}
\usepackage[normalem]{ulem}
\usepackage{xcolor}
\usepackage[hang,small,bf]{caption}
\usepackage{lineno}

\usepackage{caption}  

\newcommand{\br}{{\mathbf{r}}}
\newcommand{\bq}{{\mathbf{q}}}
\newcommand{\be}{{\mathbf{e}}}

\newcommand{\fR}{f^{\rm R}}
\newcommand{\fI}{f^{\rm I}}

\newcommand{\zetatr}{{\zeta_{\rm tr}}}
\newcommand{\zetarot}{{\zeta_{\rm rot}}}

\newcommand{\Delete}[1]{}

\newcommand{\Note}[1]{}

\begin{document}

\title{Boltzmann-Ginzburg-Landau theory for autochemotaxis of 
active rod-like particles}

\author{Shun Sakurai}
\author{Nariya Uchida}%
 \email{uchida@cmpt.phys.tohoku.ac.jp}
\affiliation{%
 Department of Physics, Tohoku University, Sendai 980-8578, Japan
}%

\date{\today}

\begin{abstract}
We investigate the interplay between chemotaxis and {alignment interactions} in active rod-like particles, such as {\it E. coli} and Janus rods. Starting from a discrete model of self-propelled rods with chemotactic responses, we employ a Boltzmann-Ginzburg-Landau (BGL) approach to derive coarse-grained dynamical equations for the density, polar and nematic orientational order parameters, and the concentration field of the chemoattractant. We perform a linear stability analysis for fluctuations around uniform steady states corresponding to isotropic and nematic phases. In both phases, we find that translational chemotactic response promotes instability, while rotational chemotactic response suppresses it, elucidating their contrasting effects on the onset of collective dynamics.
\end{abstract}


\maketitle


\section{\label{sec:introduction}Introduction}

Chemotaxis is a key mechanism governing the motion of microorganisms. Eukaryotic cells migrate toward regions of higher chemical concentration by sensing spatial gradients of attractants~\cite{parent1999cell}. In contrast, bacteria detect temporal changes in concentration and adjust their tumbling probability during run-and-tumble motion, allowing them to bias their movement toward regions of higher concentration~\cite{berg1975chemotaxis}. 
It is known that bacteria can release chemical attractants themselves and attract others via autochemotaxis, thereby forming clusters with diverse morphologies~\cite{budrene1991complex,budrene1995dynamics,mittal2003motility,berg2004coli}.
The formation of such clusters has been theoretically modeled using reaction-diffusion equations, incorporating variables such as bacterial density and the concentration field of chemical attractants~\cite{aotani2010model,brenner1998physical,keller1970initiation}.

On the other hand, many bacterial species possess rod-like shapes and exhibit collective behavior characterized by orientational alignment. Rod-shaped particles with such alignment interactions are known as self-propelled rods~\cite{ginelli2010large}. The self-propelled rod model is a modification of the classical Vicsek model~\cite{vicsek1995novel,chate2008collective},
incorporating alignment interactions in which two particles tend to align parallel if their relative angle is acute, and antiparallel if it is obtuse. Recent experiments using elongated mutant strains of bacteria ~\cite{nishiguchi2017long} 
have demonstrated that bacterial populations can exhibit long-range orientational 
order and reproduce the statistical properties predicted by self-propelled rod models.

In addition to biological systems, synthetic microswimmers such as Janus particles provide experimentally controllable platforms to study chemotactic behavior.
Janus particles have two distinct faces, one of which selectively catalyzes chemical reactions to generate concentration gradients. 
These gradients induce fluid flows that drive their 
self-propulsion~\cite{howse2007self,walther2013janus}. 
Rod-shaped Janus particles, or Janus rods~\cite{paxton2004catalytic,paxton2005motility}, 
are known to form locally aligned clusters~\cite{vutukuri2016dynamic}.

Several theoretical models have been proposed to describe chemotaxis 
at the particle level. 
These include active Brownian particle models~\cite{pohl2014dynamic,stark2018artificial}, 
{models incorporating catalytic reactions in Janus particles~\cite{saha2014clusters,tucci2024nonreciprocal},
rotational chemotaxis in chemically interacting rotors~\cite{liebchen2016pattern}, 
and a phase field approach~\cite{paspunurwar2024dynamic}.
These models predict a variety of self-organized patterns, including dynamic 
clusters, bands, and asters.
A comprehensive review of synthetic chemotaxis and collective behavior in active matter is provided in~\cite{liebchen2018synthetic}.}
Hydrodynamic interactions have additionally been incorporated into 
Fokker-Planck-type equations to predict band-like aggregation patterns~\cite{lushi2012collective,lushi2018nonlinear}.

However, to date, no theoretical model has incorporated both alignment interactions and chemotaxis.
While chemotaxis and orientational alignment have each been widely investigated, they operate through fundamentally different mechanisms: chemotaxis involves particle response to scalar concentration gradients, whereas alignment interactions arise from vectorial coupling between particle orientations. As such, their combined effects cannot be captured by simply superimposing the known results from each mechanism. A theoretical framework that treats these interactions in a unified manner is thus essential for understanding systems where both effects are present—such as in populations of rod-shaped bacteria or Janus rods that produce and respond to chemical cues while aligning locally.

In this study, we incoporate chemotaxis into the self-propelled rod model to describe the emergence of orientational order in rod-shaped bacteria and Janus rods.
We analyze the resulting collective behavior using a Boltzmann-Ginzburg-Landau (BGL) approach.
{The BGL approach~\cite{bertin2006boltzmann,bertin2009hydrodynamic,peshkov2014boltzmann} is based on the assumption of binary alignment interactions. The Boltzmann equation for the probability distribution function of particle positions and orientations generates an infinite hierarchy of equations for the moments of the orientation vector, which is truncated near a transition point to yield a coarse-grained equation of motion. This approach has been applied to systems with polar~\cite{bertin2006boltzmann,bertin2009hydrodynamic} 
and nematic~\cite{peshkov2014boltzmann} interactions, 
and is extended to chiral active matter~\cite{ventejou2021susceptibility}.
}

We consider two types of chemotactic responses of self-propelled rods that produce chemoattractants: translational chemotaxis, which accelerates the particle along the concentration gradient, and rotational chemotaxis, which reorients the particle toward the concentration gradient~\cite{pohl2014dynamic,liebchen2016pattern}.  
We derive a Boltzmann equation and truncate it at fourth order in a scaling parameter proportional to the density fluctuation near the transition point. 
Using linear stability analysis of the resulting time-dependent BGL equations, 
we examine the stability of homogeneous disordered and ordered states.

The structure of the paper is as follows. In Section II, we set up the model and derive the Boltzmann equation. In Section III, we truncate the hierarchy of equations and derive the time-dependent BGL equations. In Section IV, we conduct a linear stability analysis, and in Section V, we present the results. Finally, in Section VI, we discuss our findings in the context of related experimental and theoretical studies and conclude.

\section{\label{sec:model}Model}

We set up a two-dimensional model of self-propelled rods with chemotactic interactions mediated by the concentration field \( c(\mathbf{r}, t) \) of a chemoattractant.
We start from the discrete-time evolution equations:
\begin{align}
\mathbf{r}^{t+\Delta t}_j &= \mathbf{r}^t_j 
+ v_0\Delta t \, \mathbf{e} \left( \theta^{t}_j \right)
+ \zetatr \nabla c \left( \mathbf{r}_j^t \right) \Delta t 
\, + \, \bm{\eta}_{{\rm tr}, j}^t,
\label{eq:rt}
\\
\theta^{t+\Delta t}_j &= \arg 
\left[
\sum_{k; |\mathbf{r}_j - \mathbf{r}_k| < d_0} \mathrm{sign} 
\left\{ \cos \left( \theta^t_k - \theta^t_j \right) \right\}
 \exp \left( i\theta^t_k \right) + \eta_j^t
\right]
\nonumber\\
&\quad + 
\zetarot \, 
\partial_\theta 
\mathbf{e} \left( \theta_j^t \right) 
\cdot \nabla c \left( \mathbf{r}_j^t \right) \Delta t
+ \overline{\eta}^t_j
\label{eq:thetat}
\end{align}
where \( \mathbf{r}_j^t \) and \( \theta_j^t \) are the position and orientation of the \( j \)-th particle at time \( t \), respectively;  
\( \Delta t \) is the time step;  
\( \mathbf{e}(\theta) = (\cos\theta, \sin \theta) \) is the orientation vector;  
\( v_0 \) is the self-propulsion speed;  
\( d_0 \) is the interaction radius for alignment;  
\( \zetatr \) and \( \zetarot \) are the translational and rotational chemotactic susceptibilities, respectively;
$\bm{\eta}_{{\rm tr},j}^t$, $\eta_j^t$, and $\overline{\eta}_j^t$ are independent white noise terms. 
The translational noise $\bm{\eta}_{{\rm tr},j}^t$ is a two-dimensional white noise vector drawn from a Gaussian distribution with zero mean and standard deviation $\sigma_{\rm tr}$ for each component. 
The corresponding diffusion coefficient is given by $D_{\rm tr} = \sigma_{\rm tr}^2 \Delta t / 2$
The angular noise terms $\eta_j^t$ and  $\overline{\eta}_j^t$ are drawn 
from the Gaussian distribution  
$P(\eta) = \frac{1}{\sqrt{2\pi}\sigma} \exp\left(-\frac{\eta^2}{2\sigma^2}\right)$.

Each particle secretes chemoattractant at a rate \( a_1 \), while the substance decays at rate \( a_2 \). The chemoattractant concentration field $c(\br,t)$
evolves according to the equation:
\begin{align}
\frac{\partial c}{\partial t} = D_c\nabla^2 c + a_1 \sum_j \delta(\mathbf{r} - \mathbf{r}_j^t) - a_2 c,
\label{eq:ct}
\end{align}
where \( D_c \) is the diffusion constant.

Now we consider a continuous version of Eqs.~(\ref{eq:rt}, \ref{eq:thetat}) 
within the framework of the BGL approach.
We assume a dilute suspension of particles, such that only two particles interact at a time, 
and only upon entering each other's interaction range.
Taking the limit of small \( \Delta t \), 
we obtain a Boltzmann-type equation for the probability distribution function \( f(\mathbf{r}, \theta, t) \), 
which describes the density of particles at position \( \mathbf{r} \) with orientation \( \theta \) at time \( t \):
\begin{align}
\frac{\partial f}{\partial t}  + v_0\, \mathbf{e}_\theta
\cdot\nabla f
&= I_{\rm dif}[f] + I_{\rm col}[f]
+ D_{\rm tr} \Delta f
\nonumber \\
&
- \zetarot\, \partial_{\theta}\left(f\, \partial_{\theta}\mathbf{e}_\theta \cdot \nabla c \right)
- \zetatr\, \nabla\cdot(f\, \nabla c),
\label{eq:ft}
\end{align}
where $\be_\theta = \be(\theta)$,
\begin{align}
I_{\rm dif}[f]
&= -\lambda f(\theta)
\nonumber 
\\
&+ \lambda \int_{-\pi}^{\pi} d\theta' \, f(\theta') 
\int_{-\infty}^{\infty} d\eta \, P(\eta) 
\, \delta_{2\pi}(\theta' - \theta + \eta)
\label{eq:dif1}
\end{align}
is the self-diffusion term 
due to angular noise with tumbling rate $\lambda \sim (\Delta t)^{-1}$,
and 
\begin{align}
I_{\rm col}[f]
&= - f(\theta) \int_{-\pi}^{\pi} d\theta' \, K(\theta', \theta) f(\theta')
\nonumber \\
&\quad 
+ \int_{-\pi}^{\pi} d\theta_1 \, f(\theta_1) 
\int_{-\pi}^{\pi} d\theta_2 \, K(\theta_1, \theta_2) f(\theta_2)
\nonumber \\
&\qquad \times \int_{-\infty}^{\infty} d\eta \, P(\eta) 
\, \delta_{2\pi}(\Psi(\theta_1, \theta_2) - \theta + \eta)
\label{eq:col1}
\end{align}
represents binary interactions with collision kernel
\begin{align}
K(\theta_1, \theta_2) = 2 d_0 v_0 \left| \mathbf{e}(\theta_1) - \mathbf{e}(\theta_2) \right|
\end{align}
and nematic alignment rule
\begin{align}
\Psi(\theta_1, \theta_2) = 
\frac{1}{2} \left\{
\theta_1 + \theta_2 + \pi
\left[ H(\cos(\theta_1 - \theta_2)) - 1 \right]
\right\}.
\end{align}
Here, 
\( \delta_{2\pi}(x) = \sum_{n=-\infty}^\infty \delta(x - 2n\pi) \)
is the $2\pi$-periodic $\delta$-function, and $H(x)$ is the Heaviside step function.

We represent the distribution function using a Fourier series:
\begin{align}
f(\mathbf{r}, \theta, t) = \frac{1}{2\pi} \sum_{k=-\infty}^{\infty} f_k(\mathbf{r}, t) \, e^{-ik\theta},
\label{eq:fourier1}
\end{align}
which allows us to express the local density \( \rho(\mathbf{r}, t) \),
 the polar order parameter \( \mathbf{P}(\mathbf{r}, t) = \langle \mathbf{e}_\theta \rangle \), 
 and the nematic order parameter  
\( \mathbf{Q}(\mathbf{r}, t) = \langle \mathbf{e}_\theta \mathbf{e}_\theta - \frac{1}{2} \mathbf{1} \rangle \) 
as
\begin{align}
\rho &= f_0, 
\quad
\rho \mathbf{P} = 
\begin{pmatrix}
\fR_1 \vspace{1mm}\\
\fI_1
\end{pmatrix}, 
\quad
\rho \mathbf{Q} = \frac{1}{2}
\begin{pmatrix}
\fR_2 & \fI_2 \vspace{1mm}\\
\fI_2 & -\fR_2
\end{pmatrix},
\end{align}
where \( \fR_k = \operatorname{Re} f_k \) and \( \fI_k = \operatorname{Im} f_k \), and \( \langle \cdots \rangle \) denotes the local ensemble average.

Substituting Eq.~\eqref{eq:fourier1} into Eq.~\eqref{eq:ft}, we obtain the time evolution equation for the complex Fourier modes \( f_k \):
\begin{align}
\partial_t f_k + \frac{v_0}{2} (\nabla f_{k-1} + \nabla^* f_{k+1}) 
&= 
D_{\rm tr} \Delta f_k + 
\lambda (P_k - 1) f_k 
\nonumber \\
&\hspace{-20mm}
+ \sum_{q=-\infty}^{\infty} (P_k J_{k,q} - J_{0,q}) f_q f_{k-q}
\nonumber \\
&\hspace{-20mm}
- \frac{\zetatr}{2} \left[ (\nabla f_k) \nabla^* c + (\nabla^* f_k) \nabla c + 2 f_k \Delta c \right]
\nonumber \\
&\hspace{-20mm}
+ \frac{\zetarot}{2} \, k (f_{k+1} \nabla^* c - f_{k-1} \nabla c),
\label{eq:fkt}
\end{align}
where \( \nabla = \partial_x + i \partial_y \), \( \nabla^* = \partial_x - i \partial_y \),  
\( P_k = e^{-\sigma^2 k^2 / 2} \), and the interaction kernel \( J_{k,q} \) is given by
\begin{align}
J_{k,q} &= \frac{2 d_0 v_0}{\pi} \Bigg[
\int_{-\pi/2}^{\pi/2} d\phi \, \left| \sin \frac{\phi}{2} \right| e^{i(k/2 - q)\phi}
\nonumber
\\  
&\quad 
+ \cos\left( \frac{k\pi}{2} \right) \int_{\pi/2}^{3\pi/2} d\phi \, \left| \sin \frac{\phi}{2} \right| e^{i(k/2 - q)\phi}
\Bigg].
\end{align}
Finally, 
the chemoattractant concentration
follows the continuum version of Eq.(\ref{eq:ct}),
\begin{align}
\frac{\partial c}{\partial t} = D_c\nabla^2 c + a_1 \rho - a_2 c.
\label{eq:ctconti}
\end{align}

\section{\label{sec:truncation} Truncated dynamical equations}

Next, we truncate the hierarchy of equations in Eq.~(\ref{eq:fkt})
by considering small deviations from the uniform state:
\begin{align}
\rho &= \rho_0 \quad (\text{const.}),
\\
f_k &= f_{k,0} \quad (\text{const.}), 
\quad k \ge 1, 
\\ c &= c_0 \quad (\text{const.}). 
\end{align} 
We then apply the following scaling ansatz: 
\begin{align} \rho - \rho_0 &\sim \epsilon, 
\label{eq:scalingrho} 
\\ 
f_1 - f_{1,0} &\sim \epsilon, 
\label{eq:scalingf1} \\ 
f_2 - f_{2,0} &\sim \epsilon, 
\label{eq:scalingf2} 
\\ f_3 - f_{3,0} &\sim \epsilon^2, 
\label{eq:scalingf3} 
\\ 
f_4 - f_{4,0} &\sim \epsilon^2, 
\label{eq:scalingf4} 
\\ c - c_0 &\sim \epsilon, 
\label{eq:scalingc} 
\\ 
\nabla &\sim \epsilon, 
\label{eq:scalinggrad} 
\\ 
\partial_t &\sim \epsilon. 
\label{eq:scalingdt} 
\end{align} 
Here, \( \epsilon \ll 1 \) is a small, dimensionless expansion parameter. 
The scaling relations in Eqs.~(\ref{eq:scalingrho})--(\ref{eq:scalingdt}), 
except for Eq.~(\ref{eq:scalingc}), follow from the previous analysis of 
self-propelled rods~\cite{peshkov2014boltzmann}, 
while Eq.~(\ref{eq:scalingc}) is justified by balancing the production 
and dissociation terms in Eq.~(\ref{eq:ct}). 
The derivation of the scaling relations for self-propelled rods is summarized as follows~\cite{peshkov2014boltzmann}: 
Since the instability is driven by the nematic order parameter \( f_2 \), 
we assume \( f_2 \sim \epsilon \) near the transition point. 
Furthermore, because the velocity reversal time for self-propelled rods 
is typically much longer than the time between successive collisions, 
a propagative scaling \( \partial_t \sim \nabla \sim \epsilon^\alpha \) is assumed. 
In the equation for \( f_4 \), the dominant balance is between linear terms 
and the quadratic term \( f_2^2 \), which implies \( f_4 \sim \epsilon^2 \). 
In the equation for \( f_2 \), the balance between the diffusion term 
\( \Delta f_2 \sim \epsilon^{1+\alpha} \) and the nonlinear term \( f_2^* f_4 \sim \epsilon^3 \) 
fixes the exponent as \( \alpha = 1 \). 
In the same equation, the balance between \( \Delta f_2 \) 
and the quadratic term \( f_1^2 \sim \epsilon^2 \) confirms that \( f_1 \sim \epsilon \). 
Finally, in the equation for \( f_3 \), the gradient coupling to \( f_2 \) implies \( f_3 \sim \epsilon^2 \).

In order to fully incorporate chemotactic effects, 
we expand the equations for \( k = 0, 1, 2 \) up to \( O(\epsilon^4) \),
which is one order higher than that required for simple self-propelled rods~\cite{peshkov2014boltzmann}.
This yields the following set of equations:
\begin{align}
\frac{\partial \rho}{\partial t} + v_0 \operatorname{Re}(\nabla^* f_1) 
&= 
D_{\rm tr} \Delta \rho
\nonumber\\
&\hspace{-20mm}
-
\frac{\zetatr}{2} \left[
(\nabla \rho) \nabla^* c + (\nabla^* \rho) \nabla c + 2 \rho \Delta c
\right],
\label{eq:rhot}
\end{align}
\begin{align}
\frac{\partial f_1}{\partial t} + \frac{v_0}{2} (\nabla \rho + \nabla^* f_2) 
&= \mu_1 f_1 
\, + D_{\rm tr} \Delta f_1
\nonumber \\
&\hspace{-20mm} + 
\left[ 
P_1 (J_{1,-1} + J_{1,2}) - J_{0,-1} - J_{0,2}
\right] f_1^* f_2 
\nonumber \\
&\hspace{-20mm} + 
\left[ 
P_1 (J_{1,-2} + J_{1,3}) - J_{0,-2} - J_{0,3}
\right] f_2^* f_3 
\nonumber \\
&\hspace{-20mm} 
-
\frac{\zetatr}{2}
\left[
(\nabla f_1) \nabla^* c + (\nabla^* f_1) \nabla c + 2 f_1 \Delta c
\right]
\nonumber \\
&\hspace{-20mm} 
+ 
\frac{\zetarot}{2} (\rho \nabla c - f_2 \nabla^* c),
\label{eq:f1t}
\end{align}
\begin{align}
\frac{\partial f_2}{\partial t} + \frac{v_0}{2} (\nabla f_1 + \nabla^* f_3)
&= \mu_2 f_2 
\, + D_{\rm tr} \Delta f_2
\nonumber \\
&\hspace{-20mm} + (P_2 J_{2,1} - J_{0,1}) f_1^2 
\nonumber \\
&\hspace{-20mm} + 
\left[
P_2 (J_{2,-1} + J_{2,3}) - J_{0,-1} - J_{0,3}
\right] f_1^* f_3
\nonumber \\
&\hspace{-20mm} + 
\left[
P_2 (J_{2,-2} + J_{2,4}) - J_{0,-2} - J_{0,4}
\right] f_2^* f_4
\nonumber \\
&\hspace{-20mm} 
- 
\frac{\zetatr}{2}
\left[
(\nabla f_2) \nabla^* c + (\nabla^* f_2) \nabla c + 2 f_2 \Delta c
\right]
\nonumber \\
&\hspace{-20mm} 
+ 
\zetarot (f_1 \nabla c - f_3 \nabla^* c),
\label{eq:f2t}
\end{align}
where the coefficient
\begin{align}
\mu_k = -(1 - P_k) + 
\left[ 
P_k (J_{k,k} + J_{k,0}) - (J_{0,k} + J_{0,0}) 
\right] \rho.
\label{eq:muk}
\end{align}
characterizes the linear stability of the disordered state $f_k=0$ $(k \ge 1)$
in the absence of spatial nonuniformity.

In the following, we consider the case where \( \mu_3 \) and \( \mu_4 \) are negative,
so that \( f_3 \) and \( f_4 \) are slaved to \( \rho \), \( f_1 \), and \( f_2 \)
through Eq.~(\ref{eq:fkt}).
Then, we can neglect the time derivatives of \( f_3 \) and \( f_4 \) in Eq.~(\ref{eq:fkt}),
and express them up to \( O(\epsilon^3) \),
as they appear in Eqs.~(\ref{eq:f1t}, \ref{eq:f2t}) multiplied by quantities of order \( O(\epsilon) \).
They are given by
\begin{align}
f_3 &= \frac{1}{\mu_3}
\Bigg\{
\frac{v_0}{2}\nabla f_2 -
\left[ P_3 \left( J_{3,1} + J_{3,2} \right) - J_{0,1} - J_{0,2} \right] f_1 f_2
\nonumber\\
& \hspace{20mm} + \frac{3}{2}\zetarot f_2 \nabla c
\Bigg\},
\\
f_4 &= -\frac{1}{\mu_4}(P_4 J_{4,2} - J_{0,2}) f_2^2.
\end{align}

Substituting these expressions into Eqs.~(\ref{eq:f1t}) and (\ref{eq:f2t}),
we obtain 
\begin{align}
\frac{\partial f_1}{\partial t}
= &
-\frac{v_0}{2} \left( \nabla \rho + \nabla^* f_2 \right)
+ \mu_1 f_1 {\, + D_{\rm tr} \Delta f_1}
\nonumber\\
& + \zeta f_1^* f_2 + \beta |f_2|^2 f_1 + \gamma f_2^* \nabla f_2 + \frac{3\gamma}{v_0} |f_2|^2 \nabla c
\nonumber\\
& {-} \frac{\zetatr}{2}
\left[
(\nabla f_1) \nabla^* c + (\nabla^* f_1) \nabla c + 2 f_1 \Delta c
\right]
\nonumber\\
& {
+ \frac{\zetarot}{2} \left( \rho \nabla c - f_2 \nabla^* c \right)
},
\label{eq:f1tclosed}
\end{align}
and
\begin{align}
\frac{\partial f_2}{\partial t}
= &
- \frac{v_0}{2} \nabla f_1
+ \left( \mu_2 - \xi |f_2|^2 \right) f_2
+ \left({D_{\rm tr} \, +}  \frac{v_0^2}{4|\mu_3|} \right) \Delta f_2
\nonumber\\
& + \omega f_1^2
+ \tau |f_1|^2 f_2
+ \kappa_1 f_1^* \nabla f_2
+ \kappa_2 \nabla^* (f_1 f_2)
\nonumber\\
& {-} \frac{\zetatr}{2}
\left[
(\nabla c) \nabla^* f_2 + (\nabla^* c) \nabla f_2 + 2 f_2 \Delta c
\right]
\nonumber\\
& 
+\zetarot 
\Bigg[
{
f_1 \nabla c
+ \frac{2 \kappa_2}{v_0} f_1 f_2 \nabla^* c
- \frac{3 \kappa_1}{v_0} f_1^* f_2 \nabla c
}
\nonumber\\
& {-} \frac{v_0}{4|\mu_3|}
\left\{
3 (\nabla^* f_2) \nabla c
- 2 (\nabla f_2) \nabla^* c
+ 3 f_2 \Delta c
\right\}
\Bigg].
\label{eq:f2tclosed}
\end{align}

Eqs.~(\ref{eq:rhot}),(\ref{eq:f1tclosed}), (\ref{eq:f2tclosed}), 
and (\ref{eq:ctconti})
constitute a closed system of equations for \( \rho \), \( \vb{P} \), \( \vb{Q} \), and \( c \).
The new parameters introduced in the coarse-grained equations are expressed 
in terms of the original model parameters as follows:
\begin{align}
\gamma &= \frac{v_0}{2\mu_3}
\left[
P_1 \qty(J_{1,-2} + J_{1,3}) - J_{0,-2} - J_{0,3}
\right], \\
\beta &= -\frac{2\gamma}{v_0}
\left[
P_3 \qty(J_{3,1} + J_{3,2}) - J_{0,1} - J_{0,2}
\right], \\
\nu &= P_1 \qty(J_{1,-1} + J_{1,2}) - J_{0,-1} - J_{0,2}, \\
\omega &= P_2 J_{2,1} - J_{0,1}, \\
\kappa_1 &= \frac{v_0}{2\mu_3}
\left[
P_2 \qty(J_{2,-1} + J_{2,3}) - J_{0,-1} - J_{0,3}
\right], \\
\kappa_2 &= \frac{v_0}{2\mu_3}
\left[
P_3 \qty(J_{3,1} + J_{3,2}) - J_{0,1} - J_{0,2}
\right], \\
\xi &= \frac{P_4 J_{4,2} - J_{0,2}}{\mu_4}
\left[
P_2 \qty(J_{2,-2} + J_{2,4}) - J_{0,-2} - J_{0,4}
\right], \\
\tau &= -\frac{2\kappa_1}{v_0}
\left[
P_3 \qty(J_{3,1} + J_{3,2}) - J_{0,1} - J_{0,2}
\right].
\end{align}

\section{\label{sec:stability}Linear stability analysis}

The spatially uniform and time-independent 
solutions of the coarse-grained equations 
(\ref{eq:f1tclosed}) and (\ref{eq:f2tclosed}) satisfy
\begin{align}
0 &= \mu_1 f_1 + \zeta f_1^* f_2 + \beta |f_2|^2 f_1,
\\
0 &= (\mu_2 - \xi |f_2|^2) f_2 + \omega f_1^2 + \tau |f_1|^2 f_2,
\label{eq:solution}
\end{align}
and are given by
\begin{align}
f_1 = 0, \quad f_2 = 0
\label{eq:disordered}
\end{align}
for the disordered phase ($\mu_2 < 0$), and
\begin{align}
f_1 = 0, \quad |f_2| = \sqrt{\mu_2 / \xi}
\label{eq:nematic}
\end{align}
for the nematic phase ($\mu_2 > 0$).
In the latter case, the phase $\psi_2$ of $f_2$ is arbitrary,
and we choose $\psi_2 = 0$ without loss of generality.
We consider small deviations from the spatially uniform, steady-state 
solutions, expressed as $A(\br, t) = A_0 + \delta A(\br, t)$, with
\begin{align}
\delta A &= \left(
\delta A_c \cos (\bq \cdot \br) + \delta A_s \sin (\bq \cdot \br)
\right) e^{\Gamma t},
\end{align}
for the fields $A = \rho, \fR_1, \fI_1, \fR_2, \fI_2$, and $c$.
Here, $\bq$ is the wavevector, and $\Gamma = \Gamma(\bq)$ is the complex linear growth rate.
The govening equations 
(\ref{eq:ct}, \ref{eq:rhot}, \ref{eq:f1tclosed}, \ref{eq:f2tclosed})
are linearized with respect to the 12-dimensional real-valued vector
\begin{align}
\delta \vb{A} &= \Big(
\delta \rho_c, \delta \rho_s, 
\delta \fR_{1c}, \delta \fI_{1c},
\delta \fR_{1s}, \delta \fI_{1s},
\nonumber \\
& \hspace{15mm}
\delta \fR_{2c}, \delta \fI_{2c},
\delta \fR_{2s}, \delta \fI_{2s},
\delta c_c, \delta c_s
\Big)^{\rm T},
\end{align}
and written in matrix form as
\begin{align}
\Gamma \delta \vb{A} = \vb{M} \cdot \delta \vb{A},
\end{align}
where the non-zero components of the matrix $\vb{M}$ are given by
\begin{align}
&
{M_{1,1}  = M_{2,2} = - D_{\rm tr} q^2,}
\\
& 
M_{3,3} = M_{5,5} = 
\mu_1 
{- D_{\rm tr} q^2}
+ {\nu} f_2  + \beta \left| f_2 \right|^2,
\\
& 
M_{4,4} = M_{6,6} = 
\mu_1 
{- D_{\rm tr} q^2}
- {\nu} f_2 + \beta \left| f_2 \right|^2,
\\
&
M_{7,7} = M_{9,9} = 
\mu_2 
- \left({D_{\rm tr} \, +} \frac{v_0^2}{4|\mu_3|} \right) {q^2}
- 3 \xi \left|f_2 \right|^2 
\\
&
M_{8,8} = M_{10,10} = 
\mu_2 
- \left({D_{\rm tr} \, +} \frac{v_0^2}{4|\mu_3|} \right) {q^2}
- \xi \left|f_2 \right|^2,
\\ 
& 
M_{1,5} =  -M_{2,3} = 2 M_{3,2} = -2 M_{5,1} 
= v_0 q_x,  
\\
& 
M_{1,6} = -M_{2,4} = 2 M_{4,2} = -2 M_{6,1} 
= v_0 q_y, 
\\
& 
M_{9,3} = M_{10,4} = - M_{5,7} = - M_{6,8} =
\left(\frac{v_0}{2} - \gamma f_2 \right) q_x,
\\
& 
M_{3,10} = -M_{4,9} = M_{6,7} = -M_{5,8} =
\left(\frac{v_0}{2} + \gamma f_2 \right) q_y,
\\
& 
M_{7,5} = M_{8,6} = -M_{9,3} = -M_{10,4} =
\left(\frac{v_0}{2} - \kappa_2 f_2 \right) q_x,
\\
& 
M_{8,5} = -M_{7,6} = M_{9,4} = -M_{10,3} =
\left(\frac{v_0}{2} + \kappa_2 f_2 \right) q_y,
\\
&
M_{7,1} = M_{9,2} =
\left(
\frac{\partial\mu_2}{\partial\rho}-
\frac{\partial\xi}{\partial\rho} \left| f_2 \right|^2
\right) f_2,
\\
& 
M_{1,11} = M_{2,12} = -\zetatr \rho {q^2}  
\\
&
{M_{3,12}} = -M_{5,11} = 
{-}
\frac{\zetarot}{2} 
\left( \rho - f_2 - \frac{6\gamma}{v_0}|f_2|^2 \right) q_x,
\\
&
M_{4,12} = -M_{6,11} =
{-}
\frac{\zetarot}{2} 
\left( \rho + f_2 - \frac{6\gamma}{v_0}|f_2|^2 \right) q_y,
\\
&
M_{7,11} = M_{8,11} = M_{9,12} = M_{10,12} = 
{-}
\left(\zetatr +
 \frac{3v_0 \zetarot}{4|\mu_3|}\right) f_2 {q^2},
\\
&
M_{11,1} = M_{12,2} = {a_1},
\\
&
M_{11,11} = M_{12,12} = - D_c {q^2} - a_2.
\end{align}
The eigenvalues $\Gamma_n$ and corresponding eigenvectors $\vb{u}_n$
$(n = 1, 2, \ldots, 12)$ of the matrix $\vb{M}$ are obtained numerically.
The uniform steady-state solution is linearly unstable
if at least one eigenvalue satisfies ${\rm Re}\, \Gamma_n > 0$.

{
In the numerical analysis, we introduced dimensionless parameters 
by choosing the rod length $t_0 = \lambda^{-1}$
and $l_0 = v_0 t_0$ 
as the units of time and length, respectively, 
and fixed the other parameter values as follows:
\begin{align}
{d}_0 = 0.5,\quad
{D}_c = 0.4,\quad
{a}_1 = 0.05,\quad
{a}_2 = 0.01.
\end{align}
We also set ${D}_0 = 0$ to focus on the role of angular noise.
These dimensionless parameters are used in the linear stability analysis based on Eqs.~(39)--(58), 
unless otherwise stated.  
We also examined the case without {alignment interactions}  
by setting ${d}_0 = 0$.
}

\onecolumngrid
\vspace{0mm}
\noindent
\begin{minipage}{\textwidth}
\centering
\includegraphics[scale=0.5]{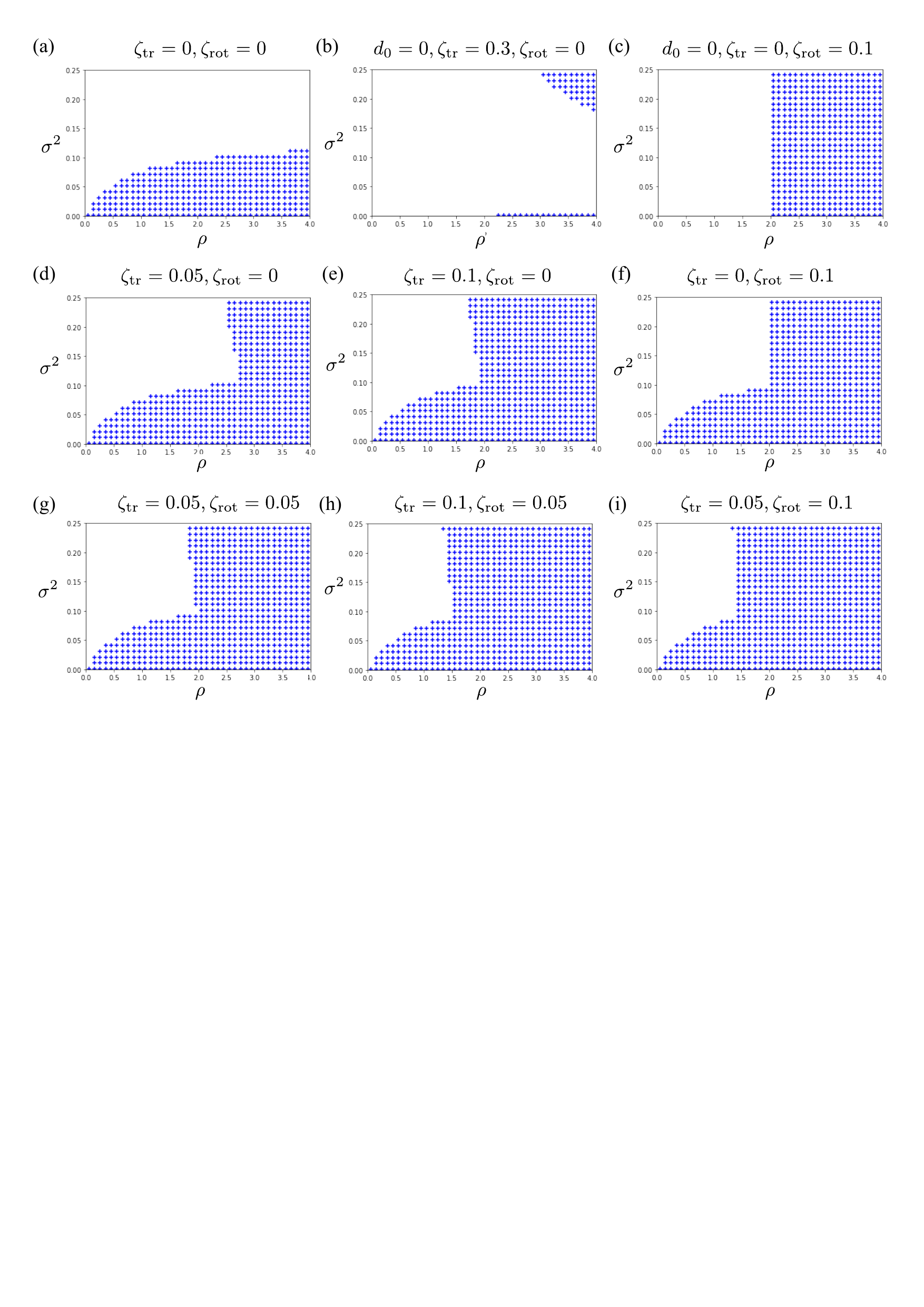}
\vspace{-100mm}
\captionof{figure}{\raggedright%
Stability diagrams for the disordered phase with different values of 
$(\zetatr, \zetarot)$:  
{(a) $(0, 0)$;  
(b) $(0.3, 0)$ with ${d}_0 = 0$, 
(c) $(0, 0.1)$ with ${d}_0 = 0$,  
(d) $(0.05, 0)$,
(e) $(0.1, 0)$,  
(f) $(0, 0.1)$,  
(g) $(0.05, 0.05)$,  
(h) $(0.1, 0.05)$, 
(i) $(0.05, 0.1)$. 
}
See the text for details.  
}\label{fig:disordered}
\end{minipage}
\vspace{1em}
\twocolumngrid 

\section{\label{sec:results} Results}

In the following,  
we present the results of the linear stability analysis using phase diagrams  
in the $\rho$--$\sigma^2$ plane,  
following the previous work~\cite{peshkov2014boltzmann}.  
Unstable states are indicated by dots in the diagrams.
\vspace{-6mm}

\subsection{\label{subsec:disordered}Disordered phase}

First, we present the results for the disordered phase.  
Figure~\ref{fig:disordered} shows the instability region  
in the $\rho$--$\sigma^2$ parameter space  
for different values of $\zetatr$ and $\zetarot$.  
The stability diagram in Fig.~\ref{fig:disordered}(a),  
where chemotaxis is absent, indicates that  
alignment interactions induces instability at  
high density or low noise levels.  
This result reproduces the previous findings for  
self-propelled rods~\cite{peshkov2014boltzmann}.
On the other hand, in the absence of {alignment interactions}  (${d}_0 = 0$),  
the translational chemotactic coupling $\zetatr$ induces  
instability primarily in the high-density and strong-noise region  
[Fig.~\ref{fig:disordered}(b)], 
{whereas the rotational coupling $\zetarot$ induces 
an instability in the high-density region,
irrespectve of the noise intensity [Fig.~\ref{fig:disordered}(c)].
}
Surprisingly, in Fig.~\ref{fig:disordered}(b), increasing the noise level—typically regarded as a stabilizing factor—leads instead 
to greater instability. 
This counterintuitive tendency can be interpreted by considering two groups of particles:
those moving in the direction of the chemical concentration gradient (parallel group) and those moving in the opposite direction (anti-parallel group).
The angular noise disturbs the motion of the parallel group and thus weakens their accumulation near the source.
However, it enables the anti-parallel group to reorient and align with the gradient direction.
If the latter effect dominates, noise can enhance the overall accumulation and thus promote instability.
{By contrast, in Fig.~\ref{fig:disordered}(c), the rotational chemotactic coupling directly induces reorientation of particles toward the concentration gradient, thereby promoting local accumulation irrespective noise-induced rotation.
}

Taking these three as the reference states, we consider the interplay of  
the orientational and chemotactic interactions.  
As shown in Fig.~\ref{fig:disordered}(d), translational chemotaxis and {alignment interactions}   
{synergistically}
enhance instability, expanding the unstable region beyond a simple superposition  
of the unstable regions in Fig.~\ref{fig:disordered}(a) and (b).  
A further increase in $\zetatr$ widens the instability region toward lower densities,  
as shown in Fig.~\ref{fig:disordered}(e).
{This synergetic effect can again be interpreted
 in terms of the two groups of particles introduced above.
The alignment interactions suppress angular fluctuations of the parallel group caused by noise,
while it may promote collective reorientation of the anti-parallel group toward the gradient direction.
As a result, both effects reinforce the accumulation of particles along the gradient, leading to enhanced instability.
}

{By contrast, the rotational chemotactic coupling and {alignment interactions} 
act independently; as shown in Fig.~\ref{fig:disordered}(f), 
the unstable region is simply the superposition of 
the regions induced by each mechanism [Fig.~\ref{fig:disordered}(a), (c)]. 
}

{The translational and rotational couplings act cooperatively;
when both $\zetatr$ and $\zetarot$ are present, 
the unstable region expands beyond the simple superposition
of the regions produced by each coupling alone,
as by comparing Fig.~\ref{fig:disordered}(i) with  (d) and (f).
Furthermore, Figs.~\ref{fig:disordered}(g)--(i) 
show that the inclination of the instability boundary depends on the ratio 
$\zeta_{\mathrm{tr}}/\zeta_{\mathrm{rot}}$; 
as $\zeta_{\mathrm{rot}}$ increases, 
the boundary becomes progressively steeper, approaching vertical.
}

\onecolumngrid
\vspace{1em}
\noindent
\begin{minipage}{\textwidth}
\centering
\includegraphics[scale=0.42]{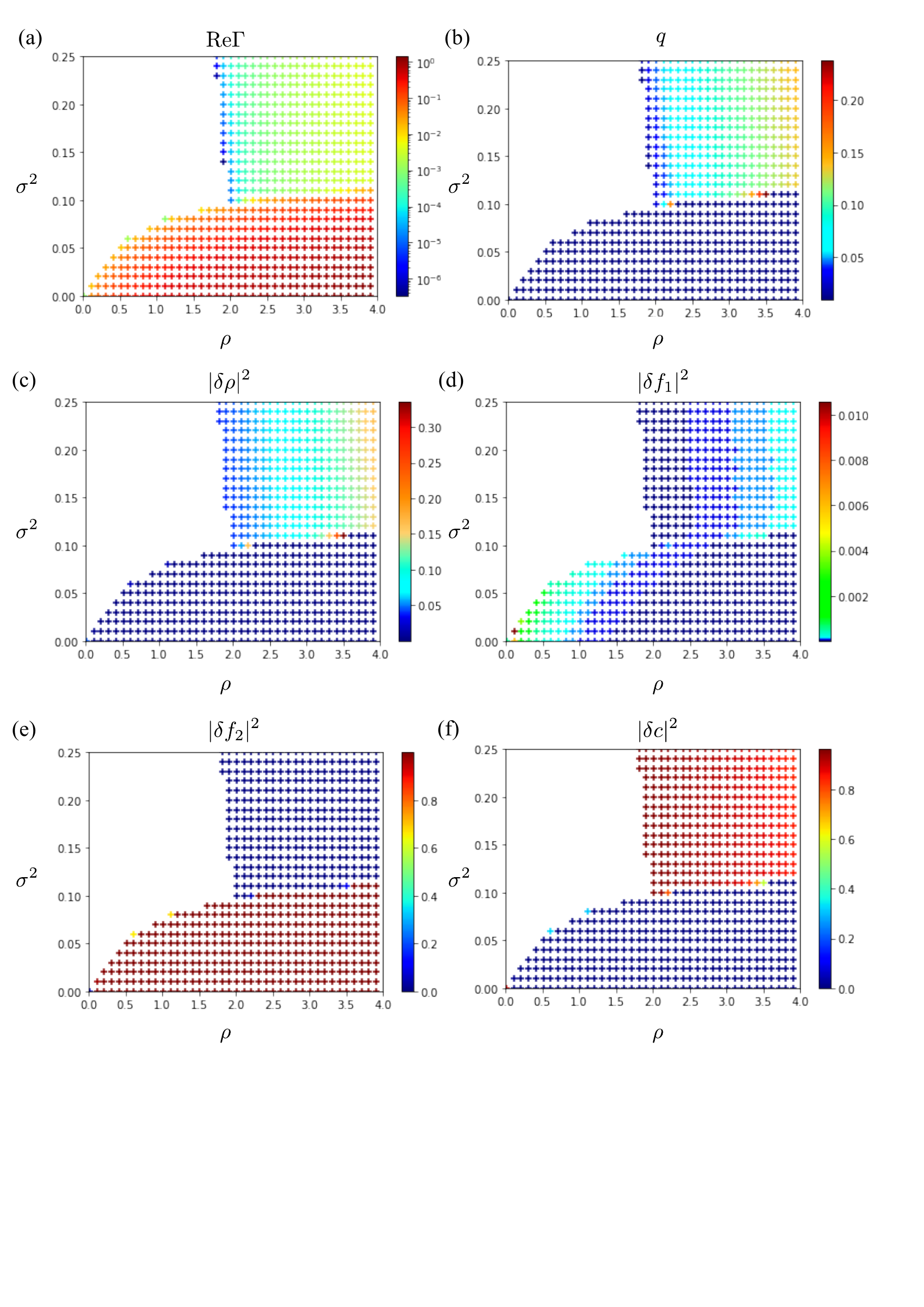}
\vspace{-35mm}
\captionof{figure}{\raggedright%
Stability diagram of the disordered phase 
for 
{$(\zetatr,\zetarot) = (0.05, 0.05)$.}
(a) The linear growth rate $\mathrm{Re}\,\Gamma(\mathbf{q})$ and 
(b) wavenumber $|\mathbf{q}|$ of the most unstable mode. 
The unstable region is shown in colorscale, and the stable region remains blank. 
(c)-(f) Sum of squared components of the most unstable eigenmode: 
(c) $|\delta \rho|^2$, 
(d) $|\delta f_1|^2$, 
(e) $|\delta f_2|^2$, 
(f) $|\delta c|^2$. 
See the text for definitions.
}\label{fig:disordered2}
\end{minipage}
\vspace{1em}
\twocolumngrid 

In Fig.~\ref{fig:disordered2}, 
we examine in more detail the stability diagram 
for $(\zetatr, \zetarot) = (0.6, 0.5)$, 
which corresponds to Fig.~\ref{fig:disordered}(h).  
Figures~\ref{fig:disordered2}(a) and \ref{fig:disordered2}(b) 
show the real part of the linear growth rate $\mathrm{Re}\,\Gamma(\mathbf{q})$ 
and the wavenumber $|\mathbf{q}|$ of the most unstable mode 
(i.e., the mode with the largest positive growth rate), respectively.
From Fig.~\ref{fig:disordered2}(a), we see that the growth rates 
in the chemotaxis-induced instability region are significantly smaller than 
those in the alignment-induced instability region [Fig.~\ref{fig:disordered}(a)]. 
This is presumably due to the fact that chemotactic coupling appears at higher order 
in the hierarchy of equations~(\ref{eq:rhot}),~(\ref{eq:f1tclosed}), and~(\ref{eq:f2tclosed}). 
Specifically, by comparing the terms involving $\zetatr$ and $\rho$ (including its derivatives) 
in Eq.~(\ref{eq:rhot}), we find that the growth rate of $\rho$ scales as $\epsilon^3$ 
according to the scaling ansatz~(\ref{eq:scalingrho})--(\ref{eq:scalingdt}): 
the first term on the right-hand side of Eq.~(\ref{eq:rhot}) is of order $\epsilon^4$, 
while the last term contains $\rho_0 \Delta c = \mathrm{O}(\epsilon^3)$. 
On the other hand, the growth rates of $f_1$ and $f_2$ are of order $\mathrm{O}(\epsilon^4)$ 
as seen from Eqs.~(\ref{eq:f1tclosed}) and~(\ref{eq:f2tclosed}).
For terms involving $\zetarot$, a similar scaling argument shows that 
the growth rate of $f_1$ is $\mathrm{O}(\epsilon^2)$. 
These rates are higher-order compared to the growth rate of $f_2$ 
in the alignment-induced instability region ($\mu_2 > 0$), 
which is dominated by the term $\mu_2 f_2 = \mathrm{O}(\epsilon)$ in Eq.~(\ref{eq:f2tclosed}). 
From Fig.~\ref{fig:disordered2}(b), 
we observe that the alignment-induced instability is most pronounced at long wavelengths, 
while the chemotaxis-induced instability is prominent at relatively short wavelengths. 
This is because the chemotactic couplings in the dynamical equations~(\ref{eq:rhot}),~(\ref{eq:f1tclosed}), and~(\ref{eq:f2tclosed}) appear only in gradient terms, 
and thus the growth rate increases with $q$ in the low-$q$ regime.


To analyze the contributions of the density $\rho$, 
polar ($f_1$) and nematic ($f_2$) order parameters, and 
concentration ($c$) to the most unstable mode,
we consider the components of the corresponding 
eigenvector $\bm{u}_{\max}$.
For example, $u_{\max,1}$ and $u_{\max,2}$ 
correspond to $\delta \rho$ 
as seen from Eq.~(39), and thus 
$|u_{\max,1}|^2 + |u_{\max,2}|^2$ 
describes the contribution of the density 
to the most unstable mode.
In Fig.~\ref{fig:disordered2}(c), we denote this quantity as $|\delta \rho|^2$ and 
show its dependence on $\rho$ and $\sigma$.
Similarly, we define the squared amplitudes 
$|\delta f_1|^2$, $|\delta f_2|^2$, and $|\delta c|^2$ 
using $\bm{u}_{\max}$ and Eq.~(39), 
and plot them in Fig.~\ref{fig:disordered2}(d), (e), and (f), respectively.
Since $\bm{u}_{\max}$ is normalized,
these squared amplitudes represent the relative contributions
of the different degrees of freedom.
The instability caused by {alignment interactions}  in low-noise regions is
predominantly governed by $f_2$. In contrast, the instability due to chemotaxis 
in high-noise regions is strongly influenced by $\rho$ and $c$. 
As the density increases, the contribution of $\rho$ becomes more pronounced, 
while that of $c$ diminishes. 
This is likely because a higher density enhances the amplitude of $\delta \rho$ 
through the term $2\rho \Delta c$ on the right-hand side of Eq.~(\ref{eq:rhot}), 
which in turn reduces the relative amplitude of $\delta c$.

\subsection{\label{subsec:nematic} Nematic phase}

Next, we consider the stability of the orientationally ordered phase.  
The stability diagrams are shown in Fig.~\ref{fig:nematic},  
where $\zetatr$ increases from the left to the right column,  
and $\zetarot$ increases from the top to the bottom row.  
The unstable regions in the absence of chemotaxis  
consist of two branches, as seen in Fig.~\ref{fig:nematic}(a),  
consistent with previous observations~\cite{peshkov2014boltzmann}.
{We refer to the upper and lower branches as regions A and B, respectively.}

\onecolumngrid
\vspace{1em}
\noindent
\begin{minipage}{\textwidth}
\centering
\includegraphics[scale=0.5]{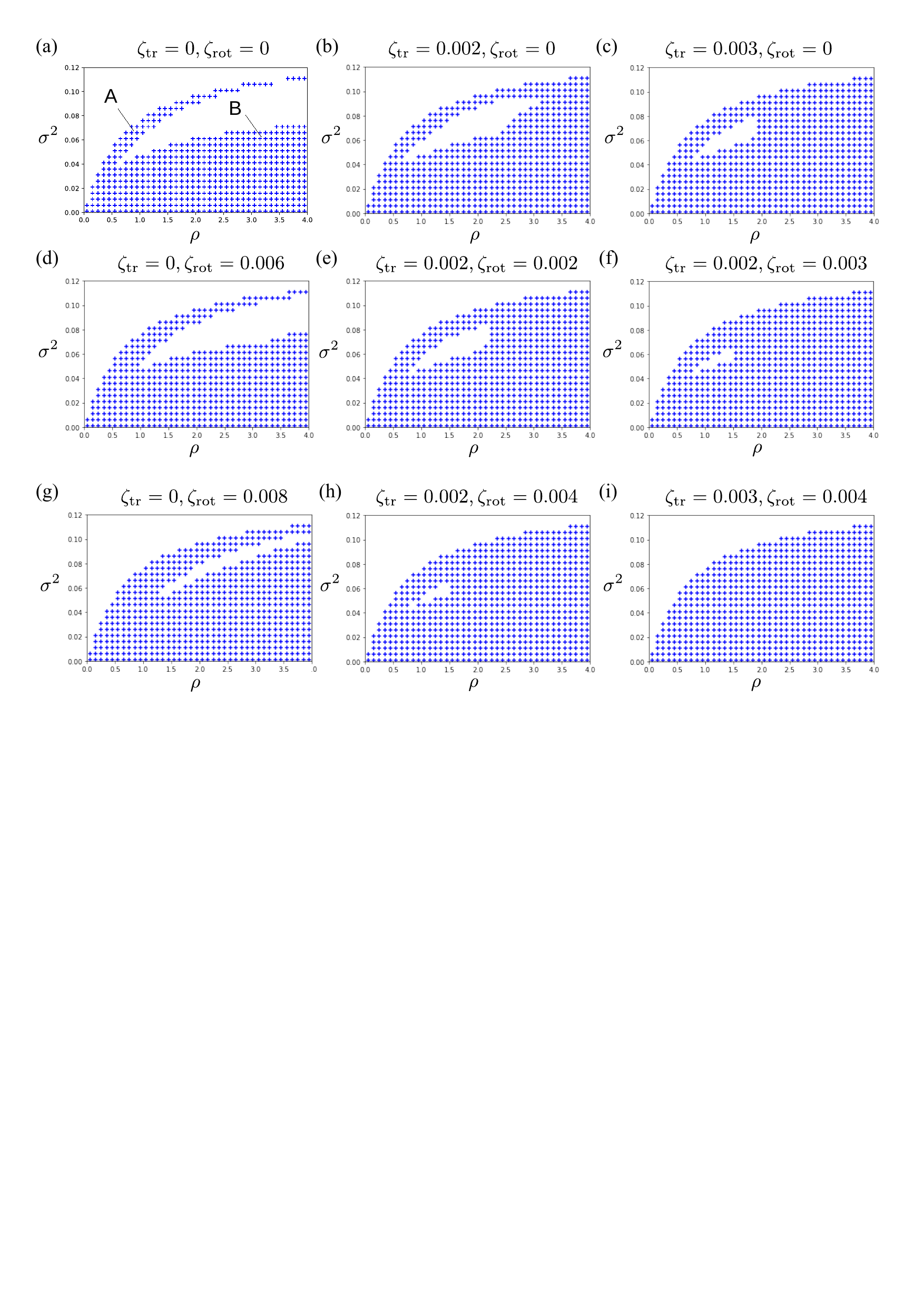}
\vspace{-100mm}
\captionof{figure}{\raggedright%
Stability diagrams for the nematic phase
with different values of 
$(\zetatr, \zetarot)$:
{(a) (0,0), 
(b) (0.002, 0),
(c) (0.003,0),
(d) (0,0.006), 
(e) (0.002,0.002), 
(f) (0.002,0.003), 
(g) (0,0.008), 
(h) (0.002,0.004), 
and
(i) (0.003,0.004).
}
{In (a), the two branches of the instability region are labeled 
as regions A and B.}
See the text for details.
}\label{fig:nematic}
\end{minipage}
\vspace{1em}
\newpage

\twocolumngrid 

{As $\zeta_{\mathrm{tr}}$ increases, the stable region between the two branches 
narrows first on the high-density side and is eventually enclosed by the unstable region, 
as shown in Fig.~\ref{fig:nematic}(b).
}
The stability ``island'' disappears with 
further increase of $\zetatr$, as seen in Fig.~\ref{fig:nematic}(c).
Note that the unstable region does not expand beyond 
the upper boundary of 
{region A} in Fig.~\ref{fig:nematic}(a); 
the alignment-induced instability remains unchanged 
{in this respect.}
This is because the upper boundary of region A is determined 
by the existence condition of the nematic phase.  
A similar tendency is observed in the transitions from (d) to (e) to (f),  
and from (g) to (h) to (i).

{
By contrast, Figs.~\ref{fig:nematic}(d) and (g) show that, as $\zeta_{\mathrm{rot}}$ increases, the unstable regions expand 
from both regions A (toward smaller $\sigma$) and B (toward larger $\sigma$). 
Comparing Fig.~\ref{fig:nematic}(e) with (b) and (d), 
we find that $\zeta_{\mathrm{tr}}$ and $\zeta_{\mathrm{rot}}$ cooperatively 
widen the unstable region, just as in the disordered phase.
}

\onecolumngrid
\vspace{2em}
\noindent
\begin{minipage}{\textwidth}
\centering
\includegraphics[scale=0.42]{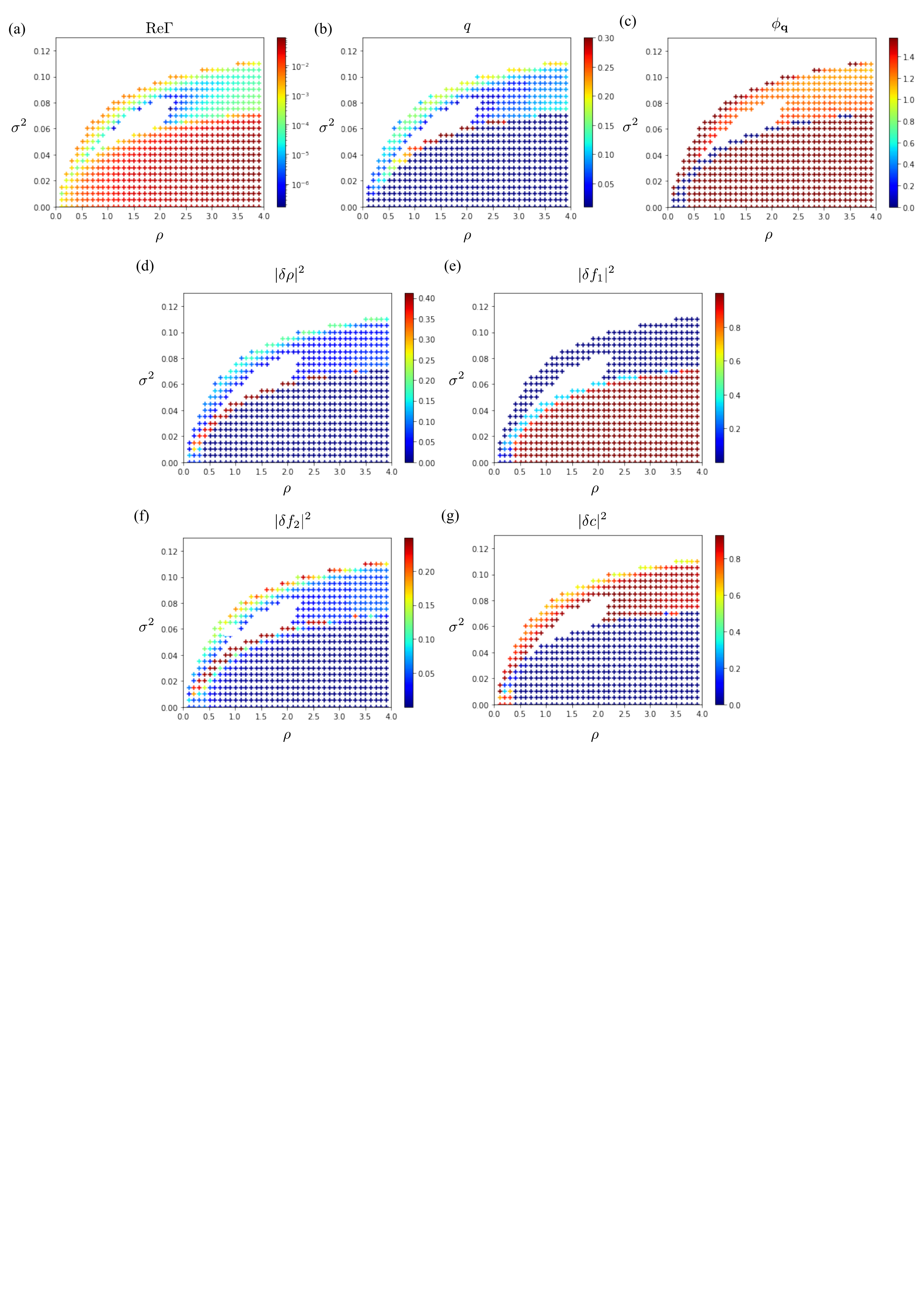}
\vspace{-110mm}
\captionof{figure}{
Stability diagram of the nematic phase for 
{$(\zetatr, \zetarot) = (0.002, 0.002)$.}
(a) Growth rate $\mathrm{Re}\, \Gamma(\mathbf{q})$, 
(b) wavenumber $|\mathbf{q}|$, and 
(c) angle $\phi_{\mathbf{q}}$ of the most unstable mode.
Unstable points are shown in colorscale. 
The unstable points are plotted in colorscale. 
The upper blank area has no nematic solution, 
and the middle blank area corresponds to the stability island, where the system is linearly stable.
(d)–(g) Sum of squared components of the eigenvector for the most unstable wavenumber:
(d) density ($\rho$), 
(e) polar order ($f_1$), 
(f) chemical concentration ($c$), and 
(g) nematic order ($f_2$).
See the text for details.
}\label{fig:ordered2}
\end{minipage}
\vspace{2em}
\twocolumngrid 

In Fig.~\ref{fig:ordered2}, we analyze the stability diagram for  
$(\zetatr, \zetarot) = (0.005, 0.005)$ [see Fig.~\ref{fig:nematic}(e)] in more detail.  
We divide the unstable region into three parts: regions A and B,  
as defined in Fig.~\ref{fig:nematic}(a), and the remaining part,  
which is attributed to chemotaxis-induced instability and denoted as region C.  
Region C connects regions A and B and encloses the stability island.

Figures~\ref{fig:ordered2}(a)–(c) show the  
linear growth rate $\mathrm{Re}\, \Gamma(\mathbf{q})$,  
wavenumber $|\mathbf{q}|$, and  
wavevector angle $\phi_{\mathbf{q}}$  
of the most unstable mode, respectively  
(i.e., the mode with the largest positive growth rate).  
The angle $\phi_{\mathbf{q}}$ is measured from the average orientation direction (taken as the $x$-axis)  
and is restricted to $0 \le \phi_{\mathbf{q}} \le \pi/2$ due to the reflection symmetry about the $x$- and $y$-axes.

In Fig.~\ref{fig:ordered2}(a), we observe that the growth rates  
are largest in region A and smallest in region C.  
From Fig.~\ref{fig:ordered2}(b), it is evident that  
long-wavelength modes are most unstable in region~A,  
whereas relatively short-wavelength modes dominate in regions~B and~C.  
{
These tendencies are consistent with the structure of the hydrodynamic equations: 
alignment interactions contribute at lower order in the small parameter $\epsilon$, 
and their effects dominate at long wavelengths. 
In contrast, chemotactic couplings appear through gradient terms 
and higher-order contributions in $\epsilon$, 
leading to instabilities at finite or shorter wavelengths. 
This pattern parallels the behavior observed in the disordered phase.
}

In Fig.~\ref{fig:ordered2}(c), 
we observe that the most unstable wavevector 
is oriented perpendicular to the underlying nematic alignment 
in regions~A and~B. 
In contrast, in region~C, the most unstable wavevector is oriented at approximately $60^\circ$  
to the nematic alignment, reflecting the distinct mechanism of chemotaxis-induced instability.

In Fig.~\ref{fig:ordered2}(d)--(g), 
we show the relative magnitudes of the density, polar and nematic 
order parameters, and concentration in the most unstable 
eigenvector, as in Fig.~\ref{fig:disordered2}.
{The dominant components of the instability vary across the regions.
In region~A, the instability is primarily influenced by fluctuations in 
$\delta f_2$ and $\delta c$,  
reflecting the strong role of nematic alignment.  
In region~B, the dominant contribution comes from $\delta f_1$,  
indicating a more polar-type modulation.  
In contrast, in region~C, $\delta c$ is the primary component,  
especially enhanced near the boundary of the stability island,  
highlighting the chemotactic nature of the instability there.
These relative magnitudes change sharply across the boundaries of regions~A, B, C, and the stable zone,  
as different eigenmodes become dominant in different areas of the parameter space.  
Such mode switching and discontinuities were also reported in the absence of chemotaxis~\cite{peshkov2014boltzmann}.
}

\section{\label{sec:discussion} Discussion}

We have used the Boltzmann approach to study the stationary solutions and their linear stability 
in a system of self-propelled particles with both chemotactic 
and {alignment interactions}. 
While chemotaxis-induced clustering has been previously explored through 
numerical simulations~\cite{pohl2014dynamic, lushi2012collective, lushi2018nonlinear, stark2018artificial}, 
and a variety of population patterns have been reported based on catalytic reaction kinetics~\cite{saha2014clusters}, 
most studies focused on either chemotaxis or alignment effects in isolation.

For {alignment interactions}, nematic ordering and clustering of elongated bacteria~\cite{nishiguchi2017long} 
and Janus rods~\cite{vutukuri2016dynamic} have been observed experimentally. 
The Boltzmann framework has been successfully applied to analytically derive hydrodynamic equations 
for such rod-like active particles~\cite{peshkov2014boltzmann}. 
Furthermore, recent works have extended the study of chemotactic mixtures to multicomponent systems 
via simulations~\cite{liebchen2018synthetic} and analytical models~\cite{tucci2024nonreciprocal, bartnick2016emerging}.

In contrast, theoretical studies combining both chemotaxis and {alignment interactions}  
have remained limited. 
In this work, we extended the Boltzmann approach of Ref.~\cite{peshkov2014boltzmann} 
to incorporate chemotactic effects decomposed into translational and rotational components, 
as proposed in Ref.~\cite{pohl2014dynamic}. 
This enabled an analytical investigation of the interplay between chemotaxis and alignment 
in self-driven rod-like particles.

Our analysis revealed several types of linear instabilities, including those driven by chemotaxis alone, 
those driven by alignment alone, and those resulting from their combined effects. 
The chemotaxis-induced instabilities arise in parameter regions not previously explored 
and are particularly relevant to dilute or noisy systems. 
{Moreover, the destabilizing effects of translational and rotational chemotactic couplings differ qualitatively, giving rise to a nontrivial synergy with alignment interactions.
}

These findings offer theoretical insight into experimental systems involving self-tactic 
and interacting particles, such as elongated {\it E. coli} bacteria and Janus rods swimming 
in aqueous hydrogen peroxide (H\(_2\)O\(_2\)) solutions. 
Future studies may extend the present framework to nonlinear dynamics, confinement effects, 
and multi-species systems, providing a more comprehensive understanding of pattern formation 
in active matter with coupled interactions.

{Finally, we comment on the possibility of quantitatively applying our model to real systems such as \textit{E.~coli}.
Based on experimental observations, typical estimates for \textit{E.~coli} are as follows:
the body length 
$\tilde{l}_0 = 2\tilde{d}_0 = 2.5\;\mu\mathrm{m}$~\cite{berg2004coli}, 
\Note{(berg2004coli,  p. 1. lushi2018 mentions 20-50 $\mu$m 
without reference)
}
Hereafter, we denote any parameter with units as $\tilde{A}$
to distinguish from its dimensionless form $A$.
self-propulsion speed 
$\tilde{v}_0 = {25}\;\mu\mathrm{m/s}$~\cite{saragosti2011directional},
\Note{(saragosti2011, Table 1, experiment)
}
mean tumbling frequency 
$\tilde{\lambda} = 3 \;\mathrm{s}^{-1}$~\cite{saragosti2011directional}, 
\Note{(saragosti2011, Table 1, experiment)
}
and 
number density $\tilde{\rho}_0 = 5\times10^8 \;\mathrm{cm}^{-3}$~\cite{saragosti2011directional}.
\Note{(saragosti2011, Materials and Methods, max density in experiment)
}
For a chemoattractant, 
the 
concentration~{$\tilde{c}_0 = 10^{-5} \mathrm{M} = 6 \times10^{15}\;
\mathrm{cm}^{-3}$~\cite{mesibov1972chemotaxis,salman2006solitary},}
\Note{(mesibov1972, Table 1, peak value in experiment.
salman2006 cited it in Fig. 5, simulation.
lushi2018 used $C_c = 1.5\times10^{17}\;\mathrm{cm}^{-3}$~\cite{lushi2018nonlinear} 
but this value is not found in the referred papers saragosti2010,2011).
berg2004coli (p.57) mentions 1 $\mu$M as a typical  aspartate concentration. 
}
diffusion constant 
$\tilde{D}_c = 8 \times10^{-6}\;
\mathrm{cm}^2/\mathrm{s}$~\cite{saragosti2011directional}, 
\Note{(used as a "typical" value. 
lushi2018 mistakenly(?) cited it as $5\times 10^{-6}$.
salman2006 used $8\times 10^{-5}$ in Fig. 5, simulation.)
}
production rate 
$\tilde{A}_1 = \tilde{a}_1 \rho_0/c_0 = 4\times10^5\;\mathrm{s}^{-1}$ molecules per bacterium~\cite{salman2006solitary,saragosti2011directional}.
\Note{(used in salman2006, saragosti2011. 
salman2006 cited S. Park et al., Proc. Natl. Acad. Sci. U.S.A. 100, 13910
(2003) but cannot be found in that paper.)
}
and 
degradation rate 
$a_2 = 5\times10^{-3}\;\mathrm{s}^{-1}$~\cite{salman2006solitary,saragosti2011directional}.
\Note{(used in simulations by salman2006, saragosti2011 
without reference to an experiment.)
}
\Note{
unit of length $l_0=v_0 t_0 = 25/3 \mu$m, 
time $t_0 = \lambda^{-1} = 1/3$ s.
}
Using these values, the corresponding dimensionless parameters are 
${D}_c = 3.8$ 
\Note{($= \tilde{D}_c t_0/l_0^2 = (8/3/(25/3)^2) \times 10^{-6+4\times2} = 3.84$)}, 
${d}_0 = 0.3$,
\Note{$= \tilde{d}_0/l_0 = 2.5/(25/3) = 0.3$}
${a}_1 = 1 \times 10^{-2}$,
\Note{$= \tilde{A}_1 t_0 \rho_0 /c_0 = (4/3 \times 5 /6) \times 10^{5+8-15} = 1.1\times 10^{-2}$}
and 
${a}_2 = 1.7 \times 10^{-3}$.
\Note{$=\tilde{a}_2 t_0 = 5/3 \times 10^{-3}$}
We have confirmed that the qualitative features of our results remain unchanged under these values.
However, different modeling choices exist for the chemotacting coupling, 
production and degradation of chemoattractants, sometimes using
nonlinear functions on bacterial density and concentration of 
chemical substances~\cite{aotani2010model, murray2001mathematical}. 
A quantitative comparison with experimental data remains 
an important direction for future work.
}

%
\begin{acknowledgments}
The authors thank Gilhan Kim and Hiroki Ishikawa for  performing
preliminary numerical simulations of related models.
\end{acknowledgments}
%

%

\end{document}